\documentclass[manuscript]{aastex}
\usepackage{lscape}
\usepackage[3D]{movie15}
\usepackage{hyperref}
\usepackage{makeidx}
\usepackage{multirow}
\usepackage{multicol}
\usepackage[dvipsnames,svgnames,table]{xcolor}
\usepackage{graphicx}
\usepackage{epstopdf}
\usepackage{ulem}
\usepackage{hyperref}
\usepackage{amsmath}
\usepackage{amssymb}

\makeatletter
	{\par\setlength{\parindent}{#3}
	\setlength{\leftmargin}{#1}       \setlength{\rightmargin}{#2}%
	\advance\linewidth -\leftmargin       \advance\linewidth -\rightmargin%
	\advance\@totalleftmargin\leftmargin  \@setpar{{\@@par}}%
	\parshape 1\@totalleftmargin \linewidth\ignorespaces}{\par}%
\makeatother 

\begin{document}
\shorttitle{Reconciling AGN-star formation}
\shortauthors{D. Garofalo et~al.}

\title{Reconciling AGN-star formation, the Soltan argument, and
Meier's paradox}

\author{David Garofalo$^1$,  Matthew I. Kim$^2$, Damian J. Christian$^2$, Emily Hollingworth$^3$, Aaron Lowery$^4$, Matthew Harmon$^5$}
\affil{$^1$Department of Physics, Kennesaw State University,
Marietta GA, 30060}

\affil{$^2$Department of Physics and Astronomy,  
California State University Northridge,  Northridge, CA 91330, USA}

\affil{$^3$Department of Aerospace Engineering, Georgia Institute of Technology,
Atlanta GA, 30332}

\affil{$^4$Department of Geosciences, Mississippi State University, MS, 39762}


\affil{$^5$Department of Physics, Southern Polytechnic State University, Marietta GA, 30060 }

\begin{abstract}
We provide a theoretical context for understanding the recent work of
Kalfountzou et al (2014) showing that star formation is enhanced at lower optical
luminosity in radio loud quasars. Our proposal for coupling the assumption of
collimated FRII quasar jet-induced star formation with lower accretion optical
luminosity, also explains the observed jet power peak in active galaxies at
higher redshift compared to the peak in accretion power, doing so in a way that
predicts the existence of a family of radio quiet AGN associated with rapidly
spinning supermassive black holes at low redshift, as mounting observations
suggest. The relevance of this work lies in its promise to explain the observed
cosmological evolution of accretion power, jet power, and star formation, in a
way that is both compatible with the Soltan argument and resolves the so-called
`Meier Paradox'.
\end{abstract}

\keywords{galaxies: active -- galaxies: star-formation -- galaxies: evolution -- galaxies: jets -
quasars: supermassive black holes- - X-rays:binaries }

\section{Introduction}

While the black hole scaling relations strongly point to a connection between
supermassive black holes and their host galaxies (Kormendy \& Richstone 1995;
Magorrian et al 1998; Ferrarese \& Merritt 2000; Gebhardt et al 2000; Tremaine et
al 2002; Marconi \& Hunt 2003), our sketchy appreciation of the link between
active galactic nuclei (AGN) and star formation suggests that understanding is
still lacking depth. Whereas the pure starburst origin to AGN (Terlevich \&
Melnick 1985; Terlevich et al 1992) is problematic, some observations suggest AGN
triggering and star formation are positively correlated (Bongiorno et al 2012;
Feltre, et al 2013), while at high luminosities they may (Luts et al 2008;
Bonfield et al 2011; Rafferty et al 2011; Juneau et al 2013) or may not (Page et
al 2012; Barger et al 2014) be.  Disk winds may suppress star formation so the
correlation may be negative (Barger et al 2014). In moderate power AGN the
possible link between AGN and star formation seems to disappear (Shao et al 2010;
Rosario et al 2012; Harrison et al 2012), and models addressing these issues
based on the variability timescales in AGN have been proposed (e.g. Hickox et al
2014). There is a need, however, for a global explanation of AGN evolution, which
AGN variability does not address, and this needs to be compatible with scale
invariance and observations of state transitions in black hole X-ray binaries.

From the Soltan argument (Soltan 1982) we conclude that measured black hole
masses and luminosities constrain the accretion history to a non-negligible
fraction of the Eddington limit, implying that supermassive black holes have
largely been spun up to high values (Fabian \& Iwasawa 1999; Elvis, Risaliti \&
Zamorani, 2002; Wu et al 2013; Trakhtenbrot 2014; Reynolds et al 2014).
Observations, therefore, require that we explain high spinning black holes at low
redshift associated with radio quiet AGN, in the context of a cosmological
evolution experiencing a downsizing of AGN activity constrained such that FRII
quasars peak at higher redshift compared to lower average redshift for FRI radio
galaxies accreting hot halo gas and quenching star formation (Cattaneo et al
2009; Merloni et al 2010; Best \& Heckman 2012). And while the most powerful FRII
quasars peak at about z=2, accretion appears to peak at later times at about z=1
(Barger et al 2001).

Our objective in this paper is to present a phenomenological theoretical
framework that has previously been applied to a number of issues in the physics
of extragalactic radio sources, to further explore the recent observations of
enhanced star formation in radio loud quasars at lower optical luminosity
(Kalfountzou et al 2014) in a way that is compatible with the implications of the
well known Soltan argument. Our goal is to flesh out and apply a simple idea:
Prolonged accretion spins black holes up and turns them into weak jet producers.
While that idea is not new, we show for the first time how we can begin an
exploration of the AGN-star formation connection within the paradigm in a way
that is intimately linked with a contradiction-free understanding of the
implications of the Soltan argument. In addition, we also show how the above
ideas resolve the so-called `Meier paradox', referring to a puzzle discovered by
astrophysicist David L. Meier in the observed redshift dependence of the radio
and optical/X-ray luminosity functions. As we will describe in more detail in the
appropriate section, Meier's observation involves a contradiction between the
expectation of how the AGN radio luminosity function and the AGN optical/X-ray
luminosity function behave as predicted by the standard black hole accretion
paradigm and the actual observations of these luminosity functions. Our
phenomenological model is based on a prolonged accretion scenario whereby black
holes spin up to high prograde values from random initial configurations in which
prograde or retrograde accretion is triggered in the aftermath of galaxy mergers,
a framework that has already been applied to address the radio loud/radio quiet
dichotomy, the FRI/FRII division, the nature of the Fundamental Plane, weak
versus strong inner disk reflection features, and the jet-disk connection
(Garofalo, Evans \& Sambruna 2010, Garofalo 2013a,b; Garofalo, Kim \& Christian
2014). In section 2 we describe the basic elements of our theoretical framework
that are needed to interpret the observations. In section 3 we explore the
results of Kalfountzou et al (2014) in terms of our model. In section 4 we
explore compatibility with the Soltan argument and address the `Meier Paradox'. 
In section 5 we juxtapose our semi-analytic framework with recent general
relativistic magnetohydrodynamic simulations. In section 6 we summarize and
conclude.

\section{The Gap Paradigm for black hole accretion and jet formation}

\subsection{Phenomenological description}
     The `gap paradigm' for
black holes is a scale-free, phenomenological model for the evolution of
accreting black holes (Garofalo, Evans \& Sambruna 2010). While retrograde
accretion is a fundamental aspect of the model, it is important to note that such
modes of accretion apply only to a small subset of the AGN population in the
paradigm. Figures 1 and 2 describe the model using a branching tree diagram. In
Figure 1 we capture the idea that major mergers in the paradigm are linked to the
most massive black holes, which can be either radio loud or radio quiet depending
on whether the cold gas forms an accretion disk in retrograde or prograde
configurations. In high retrograde spin states, the paradigm prescribes powerful
FRII quasars, whereas in high prograde spin regimes the model adopts the jet
suppression mechanism (Ponti et al 2012; Neilsen \& Lee 2009) implying a radio
quiet quasar.  For intermediate spins in retrograde configurations in the model,
we have less powerful FRII quasars while for intermediate prograde spin we have
an FRI quasar. Accretion imposes tight constraints on the evolution of these
classes of objects in the sense that the black hole spin evolution toward the
prograde regime is not a feature that the model can modify. While accretion
enforces a spin-up toward the prograde regime that is independent of model
prescription, the character of the accretion flow and the presence or suppression
of the jet depends on the ability of the FRII jet to heat the galactic medium.
For the powerful subclass, this is effective, and the radiative efficiency of the
disk drops earlier, transitioning the system into an ADAF while still in the
retrograde spin regime.  In other words, the system has not had enough time to
spin down.  For the less powerful FRIIs, the radiative efficiency drops later and
the system finds itself already in the prograde accreting regime when this
occurs, making it an FRI radio galaxy.  For a deeper appreciation of the
character of this evolution, we refer the reader to section 3.3 of Garofalo,
Evans \& Sambruna (2010). Eventually the accreting phase ends and a dead quasar
remains. For the FRI quasar, prolonged accretion will spin the black hole up
further to high spin values which, according to the model, turns the system into
a radio quiet quasar where no further evolution into other types of AGN can occur
before the system stops feeding and becomes a dead quasar. We emphasize that
while transition into radio quiet quasars is possible in the paradigm, transition
away from them is not.  The reasons for this are twofold: First, a radio quiet
quasar is a prograde accreting thin disk in the model, which additional accretion
will simply spin further up into the prograde regime, increasing the disk
efficiency and jet suppression mechanism, ensuring its radio quiet mode. Two, due
to the absence or weakness of the jet, the model prescribes that the state of
accretion will remain thin. In short, there are no mechanisms that in the
paradigm can push the system out of its radio quiet nature as long as there
continues to be sufficient material to accrete. This should be contrasted with
the behavior of X-ray binaries that do in fact experience accretion states that
transition from soft states into hard states. In other words, feeding from the
donor star can produce rather different outcomes for the state of accretion.
Therefore, FRII quasar phases are not only short (retrograde accretion can last
at most 8 x 10$^{6}$ years at the Eddington rate), they also only occur as
initial conditions, not later ones in the paradigm.  There is, i.e. in the
framework, no evolution into an FRII quasar from other active phases.  This
aspect of the model is crucial in understanding the prediction of different times
for the peak in the radio and X-ray/optical luminosity functions vs. redshift
explored in the next section.

At lower redshift the merger function drops (e.g. Bertone \& Conselice
2009) and the fraction of retrograde accreting black holes follows suit, giving
way to a preponderance of prograde accreting black holes at later times.  This is
captured in the branching-tree diagram of Figure 1 labelled `lower redshift',
with the size of the boxes capturing the density of such states.  The boxes
representing the FRII morphologies, in fact, are smaller, while those
representing prograde accreting black holes are larger.  Cosmic downsizing, thus,
has a direct impact on the generation of FRIIs in the paradigm, decreasing both
their numbers and those of the objects that are linked to them at lower redshift.

For black holes governed by secular processes, which appear to dominate AGN
feedback at least at redshift less than about 1 (e.g. Cisternas et al 2011;
Draper \& Ballantyne 2012), the AGN branching tree of Figure 2 applies.  The
crucial difference between Figures 1 and the top part of Figure 2 is the nature
of the feeding mechanism, mergers in the former and secular processes in the
latter.  From the perspective of the gap paradigm, depending on the spin and type
of accretion, we have LINERs, $\Gamma{}$-NLS1s and radio quiet AGN. The former
are the low mass equivalent of the FRI radio galaxies -- albeit in accretion
modes that are not as ineffective at launching disk winds as radio mode accretion
is - hence their spins can span the entire prograde regime.  $\Gamma{}$-NLS1s are
modeled as jetted objects in thin disk configurations, which requires that the
spin not exceed the threshold value for jet suppression, allowing them to live in
some intermediate spin range, making them lower mass black hole analogs of the
FRI quasars/AGN, but fed by secular processes thought to dominate dynamics in
spiral galaxies. The radio quiet AGN of Figure 2, finally, are simply the lower
mass equivalent of the radio quiet, high prograde spinning quasar/AGNs, but
fueled by secular processes as opposed to mergers. Again, note how the radio
quiet quasars/AGN do not evolve into other AGN states prior to terminating their
duty cycle. Because black hole spin evolution is a main driver of change due to
the tight dependence of disk and jet efficiency on spin, in the paradigm, these
objects are the slowest to evolve.  In fact, independent of the model, spinning
black holes up to high prograde values from zero spin at the Eddington limit,
requires more than an order of magnitude greater time than the spin-down from
high spin in retrograde configurations at the Eddington limit. In addition, once
spin reaches the maximum prograde value, no further spin change can occur and the
evolution of such a high-spinning prograde object is governed by the even slower
evolution related to its black hole mass.  Eventually, of course, the system runs
out of fuel and a dormant black hole is produced. For our understanding of how
the gap paradigm accommodates the Soltan argument, it is crucial that one
appreciate how all objects in this framework tend to die as high-spinning black
holes.

\begin{figure}
\vspace{-1.8in}
\includegraphics[width=400pt, height=300pt]{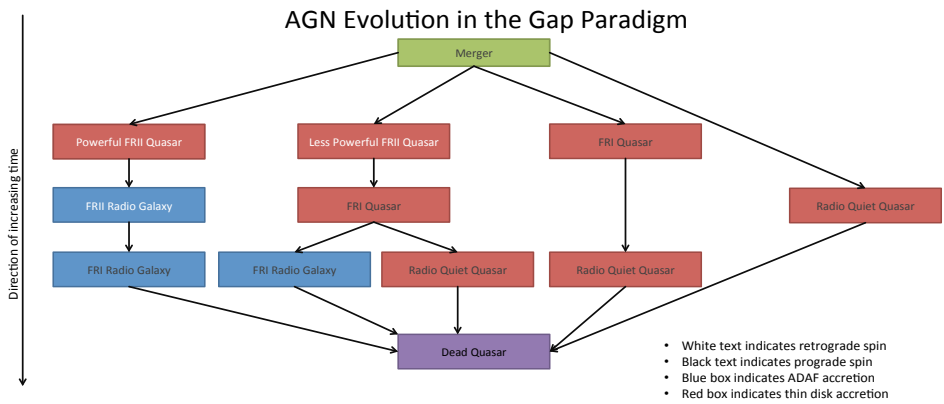}
\vspace{-0.7in}
\includegraphics[width=400pt, height=300pt]{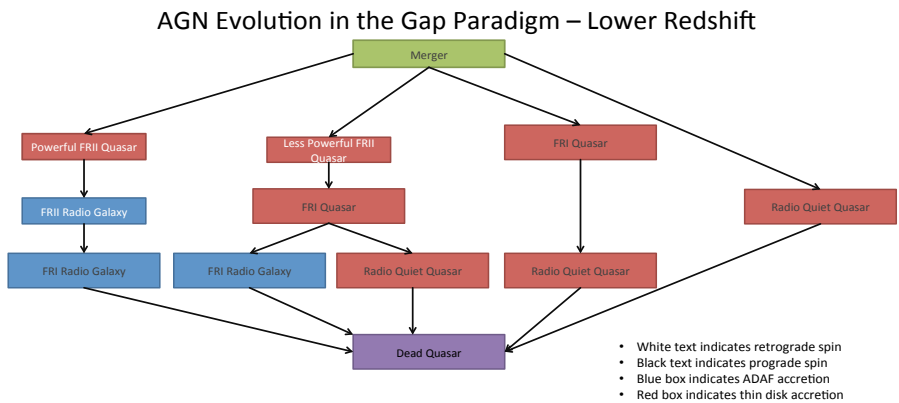}
\caption{ {\small On the top is the branching-tree diagram for the time evolution
of the most massive black holes formed in major mergers.  Two states of accretion
characterize the entire population with red representing cold, thin, radiatively
efficient accretion, and blue representing radiatively inefficient, ADAF
accretion. White labels indicate retrograde, while black labels indicate
prograde, accretion. The model predicts radio quiet quasars dominate the density
of objects at lower redshift.  On the bottom is the branching-tree diagram for
the time evolution of AGN at lower redshift at a time when the merger function
has dropped so that the number of mergers producing retrograde accreting black
holes drops.  This is captured in the diagram by the smaller sizes for the boxes
representing FRII quasars. The boxes representing prograde objects, accordingly,
are larger, capturing the fact that failed retrograde accretion states end up as
prograde ones.}
}
\label{fig1}    
\end{figure}    

\begin{figure*}   
\vspace{-1.8in}
\includegraphics[width=400pt, height=300pt]{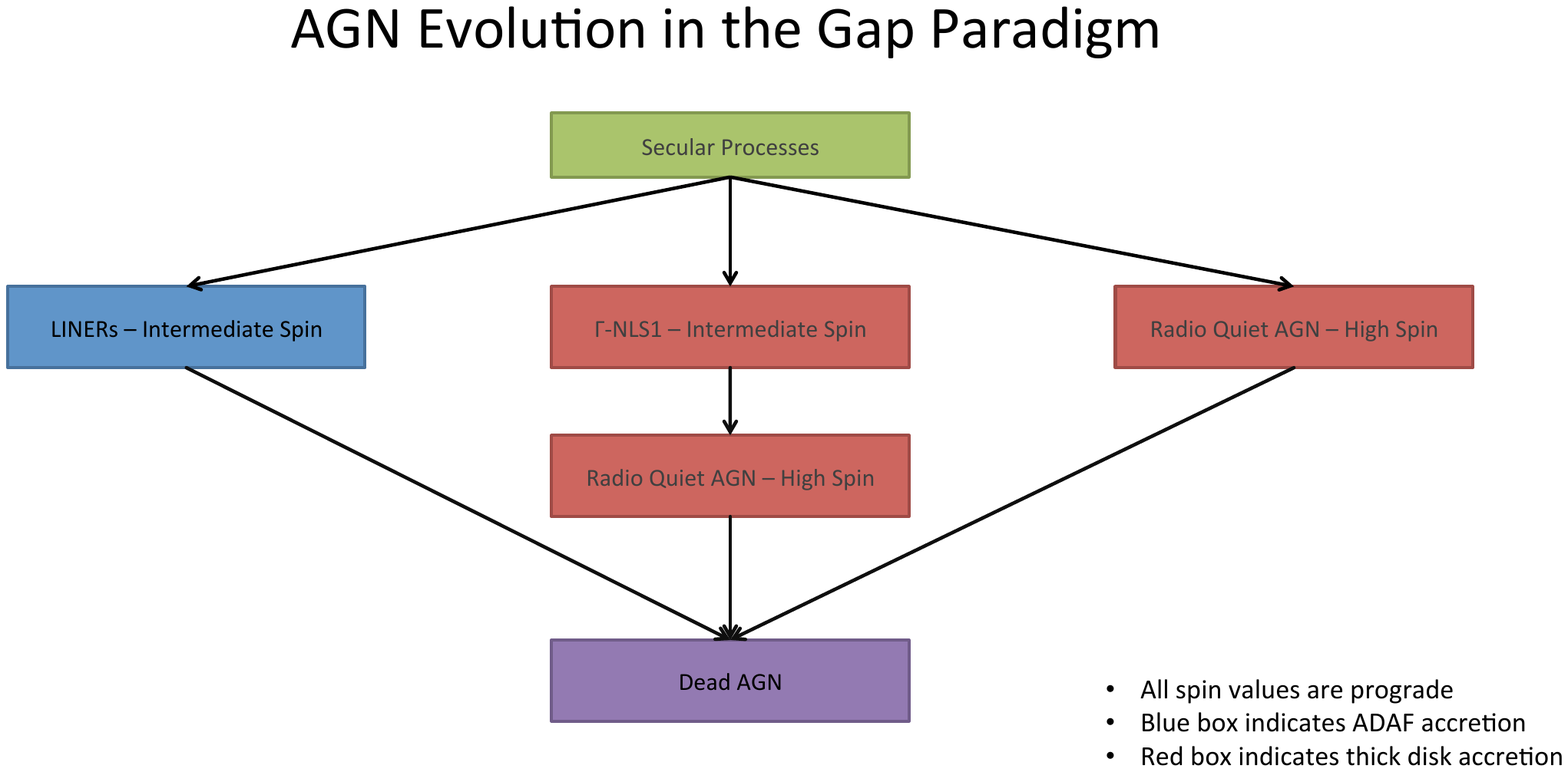}
\vspace{-0.7in}
\includegraphics[width=400pt, height=300pt]{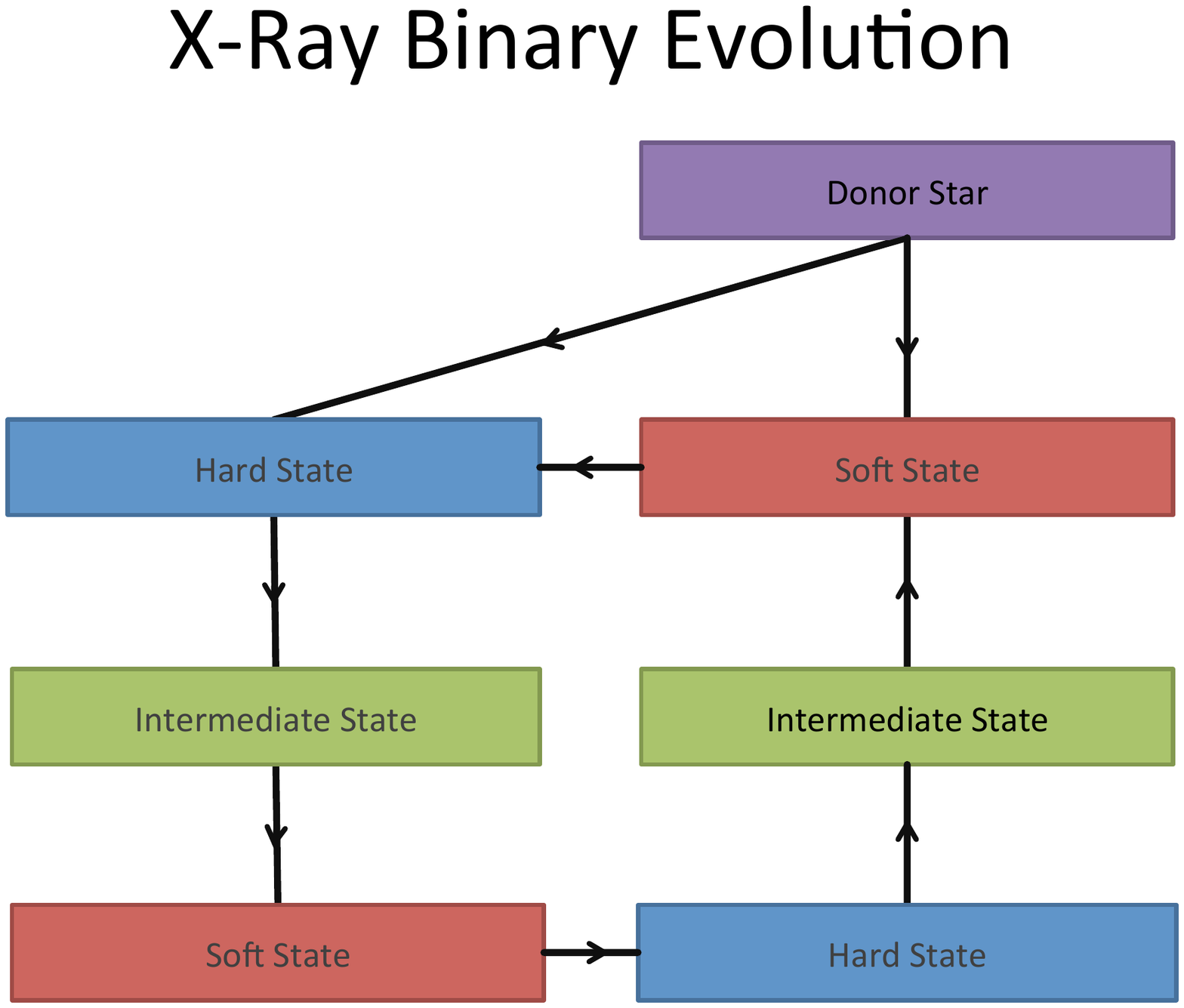}
\caption{{\small On the top is the branching-tree diagram for lower mass black
holes that are fed via secular processes.  Such lower mass accreting black holes
tend to be unstable to retrograde accretion (e.g. Perego et al 2009), effectively
ensuring the absence of FRII states.  Therefore, all black holes on this diagram
are prograde accreting. The other diagram shows the cyclical evolution of X-ray
binaries, indicating both an absence of spin evolution (black for prograde) and a
cyclical time evolution (the equal sizes of all boxes unlike the AGN case).}}
\label{fig2}    
\end{figure*}    

The time evolution of X-ray binaries, on the other hand, is insensitive to
changes in both black hole spin - which are all prograde - as well as to changes
in the distribution of the accretion states over time, as captured by the black
print and equal sizes of the boxes, respectively. But the crucial point we wish
to emphasize is that the different feeding mechanism in X-ray binaries (the donor
star) is such that soft states do\textbf{ }indeed\textbf{ }evolve into hard
states, unlike in their AGN counterparts, a fact that in the gap paradigm is
directly connected to physics that only appears in the accreting supermassive
black holes. These branching-tree diagrams will help illustrate our application
of the model to both the Kalfountzou et al (2014) work as well as to the Soltan
argument.

\subsection{Mathematical description}

The gap paradigm combines three independent theoretical constructs into one
global phenomenological model. The most fundamental involves the physics of
energy extraction from black holes via the Blandford-Znajek effect (Blandford \&
Znajek, 1977; henceforth BZ), which postulates the relation

\begin{equation}
L \propto a^2
\end{equation}

between extracted power and black hole spin parameter a. The other two
constructs involve extraction of accretion disk rotational energy via
Blandford-Payne jets (Blandford \& Payne, 1982) and accretion disk winds (Kuncic
\& Bicknell, 2004, 2007 as extensions of Shakura \& Sunyaev, 1973 and Pringle,
1981). The total outflow power from the Blandford-Znajek effect, the
Blandford-Payne mechanism and the Kuncic \& Bicknell disk wind is based on the
size of the gap region which is imposed on the following standard set of
equations. In terms of differential forms, the most concise coordinate-free way
of writing Maxwell's equations, the magnetosphere is governed by the standard
Maxwell equations with sources, which relate the exterior derivative of the dual
Faraday 2-form to the current

\begin{equation}
dF^{*}= \mu_{0} J                     
\end{equation}                                                

where
\begin{equation}
\textbf{                                          }
F^{*}= \alpha_{ij}dx_{i} \Lambda dx_{j}
\end{equation}

is the dual Faraday 2-form, \textit{$\mu{}$$_{0 }$} is the vacuum magnetic
permeability, J is the current, $\alpha{}$$_{ij }$ constitutes the tensor whose
components are differentiable electric and magnetic fields, and the summation
convention is implied. In addition, we impose the force-free condition
on the Faraday 2-form
\begin{equation}
                                              F \circ J = 0        
\end{equation}

and finally, we also impose the dissipationless ideal MHD condition

\begin{equation}
                                 F \circ U = 0         
 \end{equation}                                                           
                                                            
where U is the velocity field of the accretion flow.  In the accretion disk,
instead, the dissipationless condition does not apply and we relate the current
to Ohm's law via

\begin{equation}
dF^{*}= \sigma  F \circ U          
\end{equation}

with $\sigma{}$ the conductivity, which we treat as constant in both space and
time both for simplicity and because it is related to microscopic physics that is
not well understood. The extent to which this is a reasonable approximation is
beyond our current understanding. By comparison, general relativistic numerical
simulations of black hole accretion also impose a constant conductivity
everywhere but the value used is infinite. We will discuss this point further in
section 5. Since we work directly with the vector potential A, we have

\begin{equation}
F = dA                       
\end{equation}

so that F is the exterior derivative of the vector potential, and our equations
take the following form.

\begin{equation}
d  {\Large \wedge}  dA^{*} = \mu_{0} J
\end{equation}

\begin{equation}
dA \circ J = 0                         
\end{equation}
and

\begin{equation}
dA \circ U = 0                        
\end{equation}

in the magnetosphere, while the accretion flow is constrained by

\begin{equation}
d  {\Large \wedge} dA^{*} = \sigma dA \circ U.                                                                
\end{equation}

Stationarity and axisymmetry fully constrain the gauge. We seek solutions of

\begin{equation}
\int \frac{F}{2 \pi}  = \Psi            
\end{equation}

for a ring constructed using fixed radial and poloidal coordinates in Boyer --
Lindquist coordinates.  The \textit{$\Psi{}$} function is the invariant magnetic
flux function, i.e. the essential coordinate-free quantity whose value determines
BZ and Blandford-Payne power (Garofalo, Evans \& Sambruna 2010). On top of that
we add the power associated with the disk wind, which involves an integral over
the entire accretion disk of the local dissipation function.  This function can
be obtained from Shakura \& Sunyaev (1973) to be

\begin{equation}
D(R) = (\frac{3}{8} \pi r^{3})GM(\frac{dm}{dt})
\Big[1-({\frac{R_{isco}}{r}})^{1/2}\Big]                                  
\end{equation}

M is the mass of the black hole and dm/dt is the accretion rate. Hence, the wind
power at any location r depends on the location of the innermost stable circular
orbit, R$_{isco}$.  For locations further out in the disk, the local dissipation
from the disk will be greater in the prograde configuration due to the smaller
value of R$_{isco}$. The dependence of D(R) on R$_{isco}$ also ensures that no
stress occurs inward of that location and deviations of this have been shown to
be of order of a few percent only (e.g. Penna et al 2010). This non-relativistic
calculation is sufficient given that the relativistic correction factors drop off
rapidly with r. The bottom line is that jet and disk power depend on the location
of R$_{isco}$, which determines the size of the gap region between the edge of
the accreting material and the black hole horizon, i.e. the size of the gap
region and the imposed condition of zero magnetic flux threading the gap region
(i.e. the `Reynolds Conjecture' as coined in Garofalo, Evans \& Sambruna 2010)
constitutes the essential distinguishing feature of our model which we will
compare to numerical simulations in section 5. This zero-magnetic-flux assumption
in the gap region constitutes a simple yet fundamental difference with respect to
the BZ mechanism, which is that the gap is a region where gravity dominates the
dynamics. In the accretion disk, instead, gravity and magnetic forces compete
while everywhere else the dynamics is magnetically dominated. As described, the
equations are solved numerically to obtain the flux function, from which the
power in the jet is obtained. The details of this are discussed in the first
papers on the gap paradigm (Garofalo 2009a,b; Garofalo, Evans \& Sambruna 2010).
What concerns us here is the dependence of jet power on the spin of the black
hole, which is greater in the high retrograde spin regime (Garofalo 2009a). In
the next section we will explore the time evolution of jet power and accretion
disk power as a function of time in order to explain recent observations of star
formation in AGN.

\section{AGN-star formation link and peak in jet vs disk power}

Based on the constraints of our prolonged accretion scenario, we calculate the
redshift dependence of the luminosity of jets and accretion according to the
prescription of the gap paradigm.  If we assume a high retrograde black hole spin
as the initial configuration, the time required to spin the black hole down at
the Eddington-limit is less than about 10$^{7}$ years and about an additional
10$^{8}$ years to reach the high prograde spin regime. Due to the fact that the
gap region between the inner edge of the accretion disk and the black hole
horizon decreases in size with time, the jet power initially decreases as the
spin approaches zero but then increases although never reaching its original
strength at high retrograde spin. The opposite occurs for the disk power with the
disk efficiency increasing as the gap region decreases in size.  This behavior is
shown in the first panel of Figure 3 with blue representing the fraction of the
maximum jet power and red representing the fraction of the maximum disk power,
both as a function of redshift.  In other words, the disk power and jet power
reach their maxima at different times and the y axis captures the fraction of the
maximum power at a given time, allowing us to see when the system reaches its
maximum luminosities in jets and disks. Here, our focus is on a narrow range in
redshift in order to illustrate the basic difference in how jets and disks evolve
with time. We choose to focus on the evolution of an initially high retrograde
accreting black hole borne at redshift z=2 by following both its jet power and
disk power as a function of redshift, by taking into account the increase in mass
of the black hole associated with accretion (Moderski \& Sikora 1996). By
contrast, for objects that form or whose time evolution begins in the high
prograde accretion regime, the disk power will not display a strong redshift
evolution, as pointed out previously, due to the fact that its black hole spin
cannot change beyond its maximum value. Again, the disk luminosity can change in
this case only via a combination of decrease in accretion rate, as the
post-merger funneling of gas onto the black hole drops, and increase in black
hole mass due to accretion.  The jet power in the case of a high prograde black
hole is negligible since these accretion states correspond to radio quiet quasars
in the gap paradigm. Given this basic understanding of the physics, we can
appreciate the most glaring feature of this first panel in Figure 3, which is the
difference in the peak of accretion power vs peak jet power, with the latter
occurring at the earlier redshift. This fact survives in the paradigm regardless
of any non-zero value assumed for the initial fraction of retrograde vs prograde
black hole systems in post-mergers, due to the assumption of prolonged accretion.
If we assume radiatively efficient thin disk accretion at redshift of 2, we find
a difference in the peak between jet and disk power of about $\Delta{}$z=.07.
While jet power behaves like a damped oscillator with decrease in redshift (i.e.
it drops and then increases again but not to its original value), the disk power,
instead, steadily increases with decrease in redshift.  In particular, the two
blue and two red points near redshift of z=2 correspond to an FRII quasar jet
phase.  In the context of the work of Kalfontzou et al (2014), we see that if we
assume that jets in such objects enhance or trigger star formation (as their
observations suggest), then the model we propose naturally couples lower disk
power with larger jet power. And the optical peak in these thermal disks is
shifted to lower values for higher redshift. The accretion power varies by about
a factor of more than 20 due to the fact that the spin is evolving from
retrograde to prograde while the black hole mass increases by a factor of about 3
during that time.  We avoid drawing a continuous line because we wish to impose
Eddington-limited accretion only in an average sense. In other words, we do not
require that the system is accreting precisely at the Eddington rate at all
times, but near it, perhaps slightly above it and slightly below for periods of
time that are fractions of the total 10$^{8}$ years required to spin the black
hole up to maximal spin.

\begin{figure}   
\vspace{-1.2in}
\includegraphics[width=280pt]{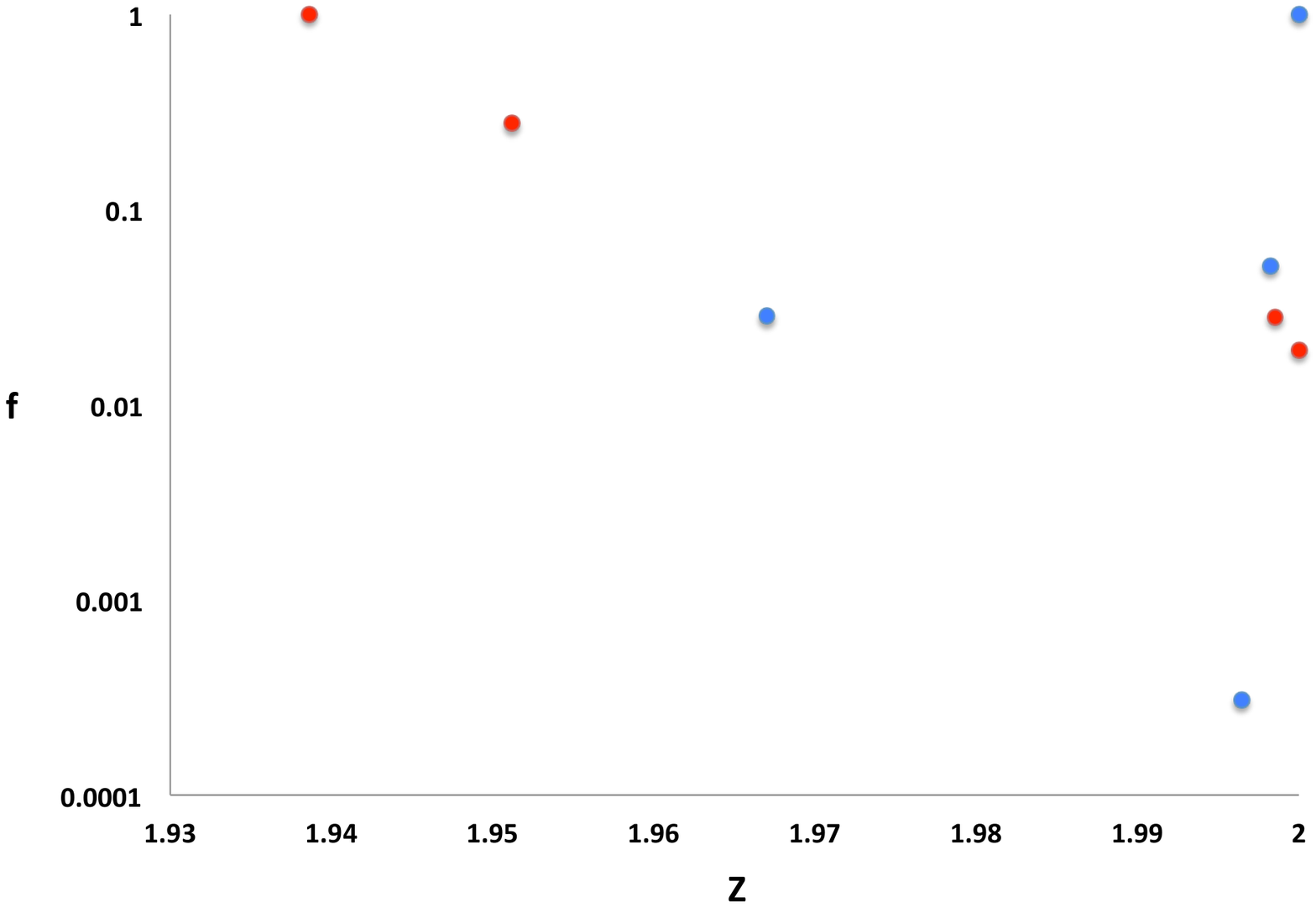}
\includegraphics[width=260pt]{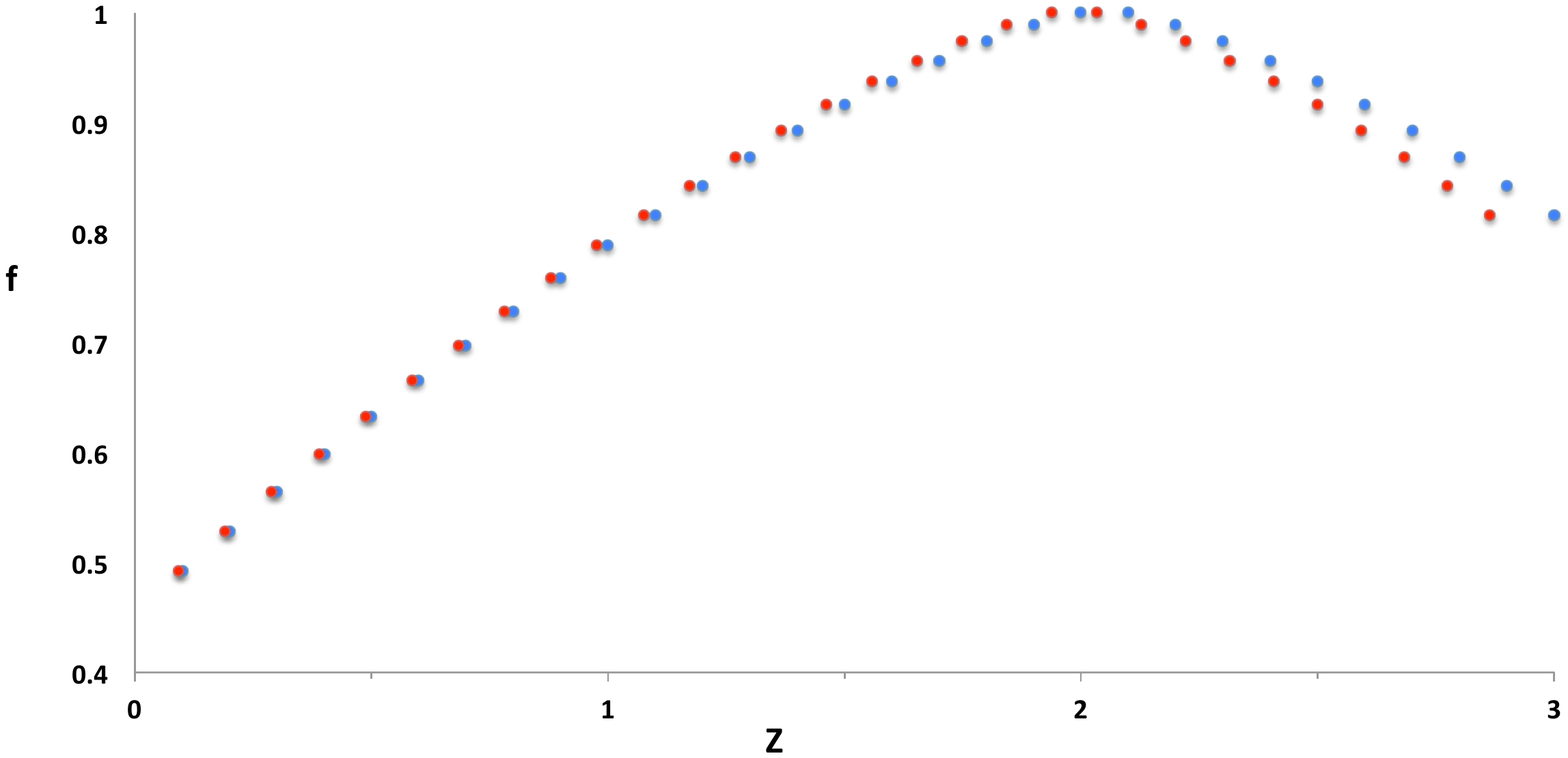} 
\vspace{-0.8in}
\includegraphics[width=400pt]{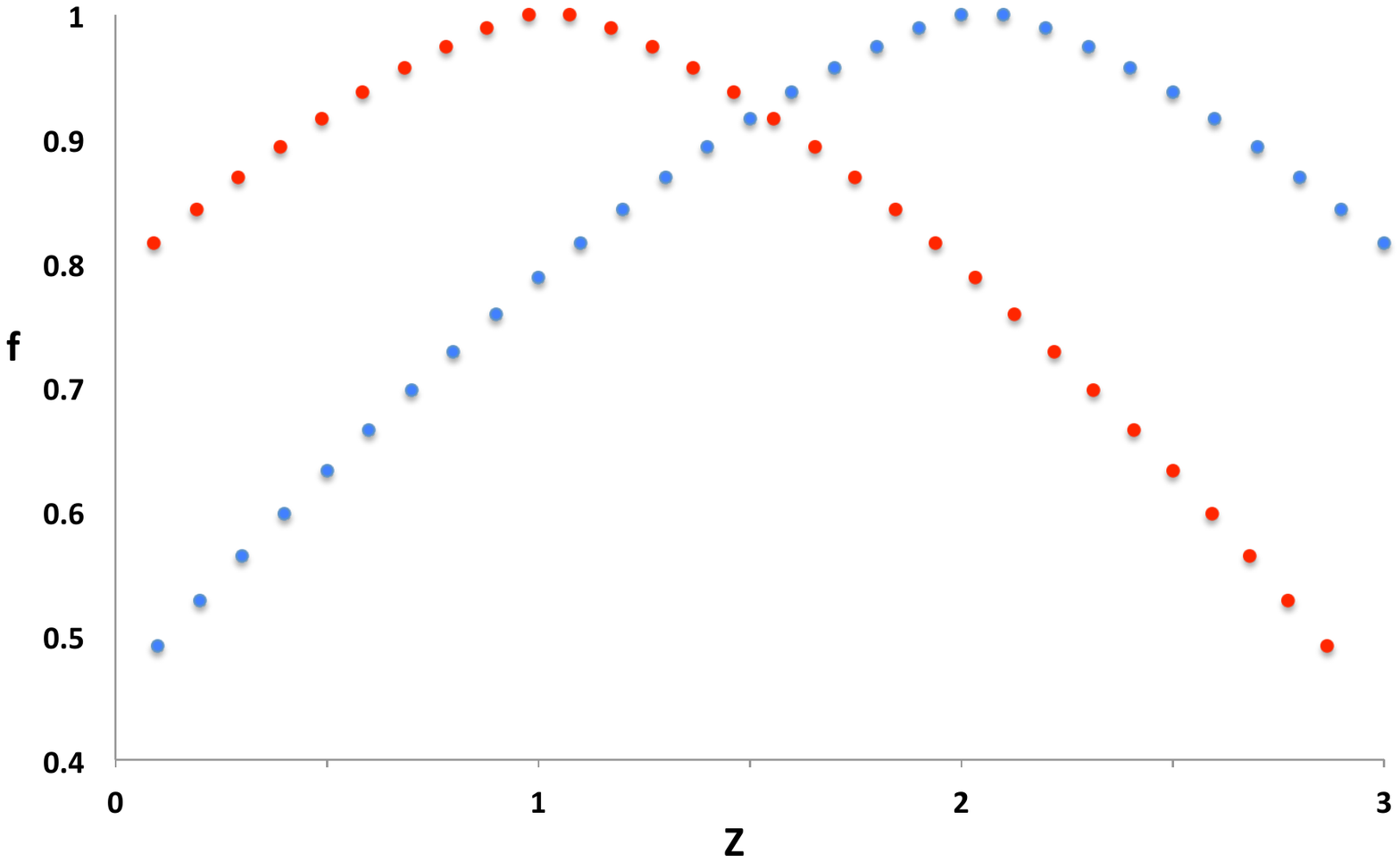}
\vspace{-0.5in}
\caption{{\small Top left: Jet and disk powers normalized to their peak
values for one accreting black hole. Blue circles represent normalized jet power
while red circles represent normalized disk power.  The maximum disk and jet
powers are both normalized to 1 for simplicity and to make it easier to see the
relative peak between the two as a function of redshift.  But there is no reason
why the maximum disk power should equal the maximum jet power. Because the disk
efficiency is lowest when the jet is most powerful, the optical peak is also
lowest. Top right: The fraction of AGN that reach maximum jet power (blue) and
maximum accretion power (red) normalized to the peak of such occurrences vs
redshift for the most massive black holes.  Using an average for the fractional
merger function from Bertone \& Conselice (2009), we have assumed that FRII
quasars are produced in equal fractions compared to radio quiet quasars, which
means that a fraction of the mergers lead to high spinning, retrograde accretion,
and an equal fraction to high spinning prograde accretion.  We see that the peak
in jet power leads the peak in accretion power at all redshifts.  Bottom:
Including lower mass accreting black holes and secular processes, shifts the disk
power peak to even lower redshift. }}
\end{figure}

In Figure 3 top right panel, instead, we explore
the full range in redshift by using the observed merger fraction for galaxies
above 10$^{9}$ solar masses (Bertone \& Conselice 2009).  There are two
differences here compared to the left panel. First, we are considering a family
of cold gas accreting supermassive black holes as opposed to one object. Two, we
are considering a larger range in redshift. Because the merger function drops as
the redshift drops below a \textit{z} of about 2, the total power in either jets
or disks will decrease with redshift below \textit{z}=2.  The peak of power on
this plot, therefore, represents the normalized sum over a family of radio loud
and radio quiet quasars of the total jet and disk powers. Accordingly, the peak
of jet power occurs at redshift of 2 on this plot due to the fact that the merger
function is highest at this redshift for the most massive black holes (Bertone \&
Conselice 2009). And, because disk power reaches its peak at about 10$^{8}$ years
after jet power, we are still capturing the same effect that appeared in the
first panel, i.e. the disk power reaching its maximum at later times compared to
jet power.  Hence, the top panel on the right captures the evolution of the total
normalized disk and jet power over a large range in redshift. We assume that
mergers lead to equal fractions of high-spinning black holes in retrograde and
prograde accretion, i.e. to equal fractions of radio loud and radio quiet
quasars.  While the radio quiet population evolves slowly, the FRII quasar class
evolves quickly, turning into radio quiet quasars in less than about 10$^{8}$
years at the Eddington accretion rate. The figure includes only the most massive
black holes (i.e. above 10$^{9}$ solar masses), explicitly done to highlight the
differences that remain even though entire families of accreting black holes are
missing. As we more realistically explore total luminosity functions by including
the contribution of lower mass black holes associated with cosmic downsizing to
this, as well as secular processes at lower redshift, we end up in the conditions
captured by the lower panel of Figure 3.  While the quantitative details depend
on particular assumptions for the importance of secular processes and mergers in
the lower mass black hole regime, the basic difference between the two peaks
appears as a noticeable shift compared to the top right figure. Because AGN
activity is shifting to lower mass accreting black holes that do not involve
retrograde accretion, radio quiet quasars/AGN are beginning to dominate the
energetics compared to the radio loud quasar/AGN group. From the perspective of
our branching-tree diagrams, AGN activity is now dynamically dominated in a way
that is captured by the second panel of Figure 1 and the first panel of Figure 2,
with fewer FRII quasars forming, and secular processes becoming more dominant. We
have added the contribution of massive black holes that are a factor of 10-100
times smaller than the ones that appear in Figure 3 top right panel. And, as we
did for the FRII quasars and radio quiet quasars with masses above 10$^{9}$ solar
masses, we use the following expression for jet power from Garofalo, Evans \&
Sambruna (2010),

\begin{equation}
L_{jet} = 2 x 10^{47} erg
s^{-1} \alpha \beta^{2} (\frac{B_{d}}{10^{5}} G)^{2} m _{9}^{2} j^{2}     
                                                           \end{equation}

where $\alpha{}$ captures the coupling between the Blandford-Znajek and
Blandford-Payne processes, $\beta{}$ prescribes the magnetic flux enhancement on
the black hole due to the gap region, B$_{d}$ represents the magnitude of the
magnetic field threading the inner accretion disk, m$_{9}$ the black hole mass in
terms of one billion solar masses, and j is the dimensionless spin parameter.
Standard thin disk power, instead, is given by the integral over the entire
accretion disk of the dissipation function D(R) given above. The difference with
this lower black hole mass population is simply in the black hole mass, which
gives smaller powers for both jets and disks. In order to capture the
observational fact that secular processes begin to dominate at redshift of 1, we
impose a peak in the contribution of 10$^{7}$ -- 10$^{8}$ accreting solar mass
black holes to disk power at about a redshift of 1. The blue function continues
to drop because the majority of the objects that are forming at lower redshift
are prograde accreting systems which can only produce jets when in the
intermediate spin range as described in the branching-tree diagrams for cold mode
accretion and these jets are less powerful compared with their retrograde cold
mode accreting counterparts.  Although the most massive quasars are no longer
contributing to the red points at lower redshift, the large numbers of less
massive accreting black holes are overwhelmingly radio quiet AGN as the redshift
approaches 1 and decreases, making the overall blue peak shift considerably from
that of the red peak. While the 1.05 difference in redshift between the peak in
accretion and jet power should not be taken too rigorously because the
uncertainties depend on rough estimates of the decrease in production of
retrograde systems, on the specific fractions of lower mass accreting black holes
contributing to the AGN phenomenon, as well as the details of the contribution of
secular processes as a function of redshift, the bottom line is the existence of
an inescapable and noticeable break between the jet and accretion peaks in the
paradigm. In other words, by working within the uncertainties in the physics we
can decrease or increase the redshift difference between the two peaks but cannot
wash away the existence of a shift\textbf{.} If the jet power is observed
predominantly in radio, and disk power in optical/X-ray, our results are
qualitatively compatible with observations (e.g. Singal et al 2013 Figure 12).
This is illustrated in the bottom panel of Figure 3. Note, finally, how the red
points representing the fractional disk power drop more slowly with decrease in
redshift. This is due to the aforementioned fact that disk-dominated objects
evolve more slowly than jet-dominated ones in the paradigm. As one decreases, and
eventually completely eliminates, the contribution of lower mass black holes and
secular processes from the bottom panel in Figure 3, again the difference becomes
miniscule and we are back in the regime described by the top right of Figure 3.
Indeed, the take away message here is that it is not the FRIIs that are
responsible for the significant lag between the difference in peak between disk
and jet powers.~ In fact, the top right of Figure 3 shows a negligible difference
between those peaks.~ It is only when you include the effect of the AGN
population as a whole that you get the large offset. However, it is important to
note that although including only the most massive accreting black hole
population produces a small difference, that small difference in peaks remains as
a result of the fundamental distinguishing feature of the paradigm: Retrograde
accretion is dynamically jet-dominated, but invariably evolves toward disk-
dominated states.

In terms of the connection with star formation, thin disk efficiency is
lowest for the highest retrograde spin and increases monotonically as the spin
becomes more prograde (Figure 4).
\begin{figure*}   
\includegraphics[width=500pt]{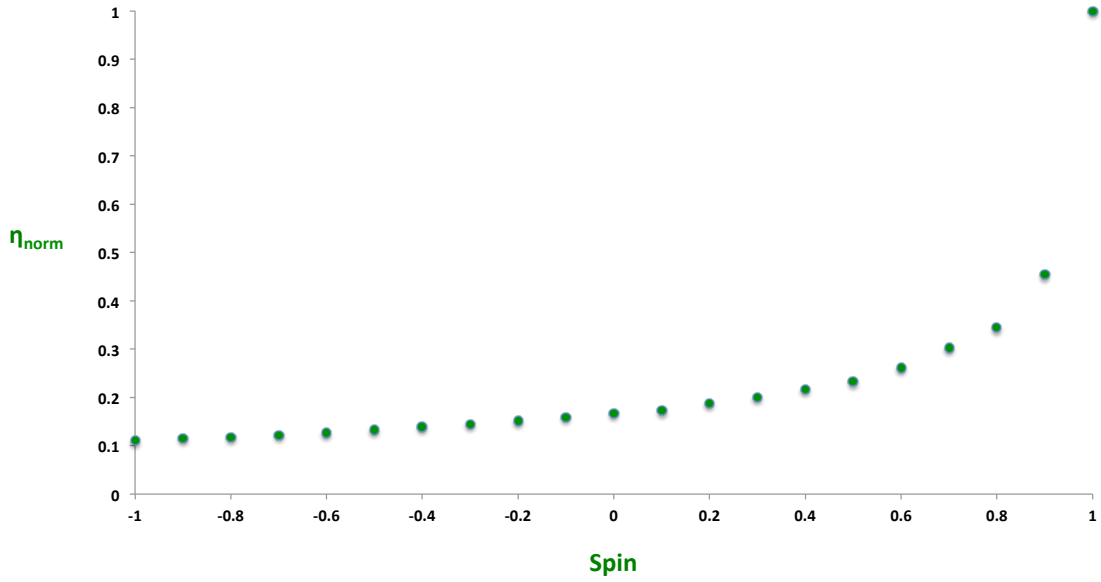}
\caption{Normalized thin accretion disk efficiency. The normalized
efficiency is in terms of the maximum possible efficiency which is at highest
prograde spin.  Because the innermost stable circular orbit is closer to the
black hole as the accretion is in greater prograde regimes, the disk efficiency
increases monotonically in the prograde direction up to 42\% of the accreted rest
mass. Negative values indicate retrograde accretion while positive values
represent the prograde regime.}
\label{fig4}
\end{figure*}

At highest retrograde accretion we have the most powerful, most collimated,
jets, which presumably induce star formation (Kalfountzou et al 2014).  But such
disks are also the least efficient (Figure 4).  At lower retrograde spins, say
around -0.2 or -0.1, the accretion efficiency has increased by 1.36 to 1.43 times
compared to highest retrograde spin. In addition, the black hole has accreted 10
to 15 percent of its original mass so it went from a mass M to a mass of up to
1.15M. The accretion power, therefore, increases both as a result of the change
in efficiency as well as to the change in black hole mass. In order to put this
on more quantitative footing, we calculate the ratio of the temperatures in the
inner disks at 10 Schwarzschild radii for disks that are the result of time
evolution from high retrograde spin of about -1 to a retrograde spin of -0.5.  We
apply standard thin disk theory and its temperature profile according to

T $\propto{}$ r $^{-3/4 }$M$^{1/4 }$[1
-- (r$_{ISCO}$/r)$^{1/2}$]$^{1/4                                                 
                    }$(15)

where M is the black hole mass, r$_{ISCO}$ is the innermost stable circular
orbit, and r is the radial location in the disk where we evaluate the temperature
(10 Schwarschild radii).  The black hole has increased its mass to about 1.1M
from the initial value M and r$_{ISCO}$ has dropped from 9 Schwarzschild radii to
about 7.5 Schwarzschild radii.   We find

{\raggedright
T$_{f}$/T$_{i}$ = 1.21  $>$ 1      
                                                           (16)
}

where T$_{i}$ is the temperature in the disk at 10 Schwarzschild radii when the
black hole is spinning at the highest retrograde value and T$_{f}$ is the
temperature in the disk at 10 Schwarzschild radii when the black hole mass has
increased by about 10\% of its original value and the spin is -0.5.  What matters
to our discussion is not the actual ratio but the fact that it is larger than
unity, a fact that is independent of the actual boundary value of the temperature
in the inner disk. As a result of this temperature increase, the peak in the
spectral intensity distribution will shift to higher energy, thereby increasing
the optical luminosity. And this is true despite the fact that 10 Schwarzschild
radii is further away from the event horizon when the black hole mass is larger.
But this increase in the energy reprocessed in the disk is associated with an
increase in disk winds (Kuncic \& Bicknell 2004, 2007) that, unlike collimated
jets, may constitute negative feedback (Barger et al 2014; Kalfountzou et al
2014). To reiterate, lower optical luminosities are associated with higher
retrograde accreting objects, precisely the class that in the paradigm produce
the most powerful, most collimated jets, by maximizing both the Blandford-Znajek
and Blandford-Payne jets (Garofalo, Evans \& Sambruna 2010).  As accretion spins
the black hole down toward zero spin and the disk efficiency and black hole mass
increase, the optical luminosity of the thin disk also increases.  However, jets
are becoming less powerful and less collimated due to the decreased size of the
gap region (Garofalo 2009b; Garofalo, Evans \& Sambruna 2010) and, therefore, are
less efficient in their ability to enhance the star formation. This picture is
qualitatively in agreement with the observation of Kalfountzou et al (2014) that
as the optical luminosity increases (they quote a threshold value of
log$_{10}$(L$_{opt}$/W) $<$ 38), the far infrared signatures of enhanced star
formation disappear. And they ask what the mechanism could be that halts the
positive feedback they assume the jets are having in the lower optical range. 
But this is expected in the gap paradigm: The optical luminosity of these
accreting black holes increases as the black holes spin down from high retrograde
values in tandem with the increase in black hole mass and disk efficiency. But,
as noted, a lower retrograde spin implies less powerful and less collimated jets.
Hence, spinning the black hole down in the retrograde regime decreases the
alleged positive feedback of jets, while increasing the alleged negative feedback
of disk winds. It is also important to emphasize that we are not appealing to
differences in accretion rates. Everything else being equal, larger accretion
rates will enhance both the jet and disk powers, which of course, have
competing/opposite effects. The distinguishing element here, instead, is the disk
efficiency, which depends on black hole spin.  Larger disk efficiency is coupled
to both less powerful and less collimated jets. It is also important to note that
the jet-driven star formation we are advocating here is not simply a matter of
larger jet power. At lower disk efficiency, the theory prescribes that both jet
power and jet collimation, are larger. And, as Kalfountzou et al (2014) point
out, the fact that the star formation rate at higher overall power appears to
remain the same as that in the lower optical regime, is compatible with the idea
that while the jet-induced star formation is greater, the competing negative
feedback due to stronger disk winds compensates, i.e. the disk efficiency is such
that the tug-of-war between jets and disks is in balance. Overall, we come to the
following explanation: Under the assumption that collimated FRII jets enhance or
trigger star formation, and that un-collimated or less collimated disk winds
produce negative feedback, cold mode accreting systems in high spinning
retrograde configurations will \textit{initially} enhance the star formation, but
inevitably evolve toward accretion states that have the opposite effect. The
conversation in terms of a tug-of-war or a competition between jets and disks
described in Garofalo, Evans \& Sambruna (2010), is transferrable to a
conversation about the star formation connection with AGNs. Accordingly, black
hole accretion borne in the far retrograde regime, will experience a shift not
only in the nature of this active phase (jet dominated to disk dominated), but
also in its ability to influence the host galaxy star formation by shifting the
nature of its feedback (from positive to negative). It goes without saying that
we have not shown that the Kalfountzou et al work cannot be interpreted in other
ways. We have described how the jet-induced star formation suggested by others
(Tadhunter et al 2014; Gaibler et al 2012; Crocket et al 2012; Croft et al 2006)
makes sense within the gap paradigm. Very much related to the time evolution of
black hole accretion used so far in this work, in the next section we include a
discussion of the Soltan argument, showing how these ideas just presented allow
one to also appreciate how massive, high-spinning black holes at low redshift,
should be both common and radio quiet.

\section{The Soltan argument: High spinning black holes
and radio quiet AGN at low redshift}

As emphasized in previous work (Garofalo, Evans \& Sambruna 2010; Garofalo
2013a,b), high-spinning, prograde accreting black holes in radiatively efficient
states, constitute the most effective conditions for the absence of jets. This
disk quenching or suppression of the jet comes from observations of X-ray
binaries (Ponti et al 2012; Neilsen \& Lee 2009), thereby ensuring scale
invariance. For post-merger, prograde, high-spin black hole accretion, jet
suppression ensures such systems behave like radio quiet AGN/quasars, but for
systems that begin in the retrograde regime, such jet suppression states are the
product of time evolution, reached after about 10$^{8}$ years at the Eddington
accretion rate. Because the model prescribes the generation of hot gas accretion
as the result of a previous FRII quasar phase that was powerful enough to heat
the galactic medium and affect accretion during the timescales of the accretion
process (Antonuccio-Delogu \& Silk 2010), most AGN will not find themselves
accreting in these hot, radiatively inefficient states, and will therefore reach
the high-spin black hole regime in radiatively efficient, radio quiet or jet-less
states. This is captured in the branching-tree diagrams of Figures 1 and 2, with
the size of the boxes representing FRI radio galaxies becoming smaller at  lower
redshift, and by being absent on the secular processes diagram (Figure 2). In
particular,~FRI~radio galaxies are explicitly forbidden in the model to emerge
from mergers so there are no arrows that connect the merger box to the FRI radio
galaxy box.~Therefore, within the paradigm, with the exception of the largest
accreting black holes (whose density is in decline), decrease in redshift ensures
that a large population of prograde accreting black holes must be both spinning
rapidly and be associated with an absence of strong jets, ensuring compatibility
with the Soltan argument. Note how cosmic downsizing implies that fewer FRII
quasars will form at low redshift since FRII quasars require retrograde accretion
in the paradigm which occurs only if the following conditions apply: 1) They
occur statistically only in a subset of major mergers but the merger function
drops with decrease in redshift (Bertone \& Conselice 2009; Treister et al 2012);
2) retrograde accretion is unstable unless the black hole is massive (Perego et
al 2009; Garofalo 2013b), so less massive black holes are less likely to live
long in retrograde configurations. And the observations in fact support this idea
(Dunlop et al 2003; McLure \& Jarvis 2004; Floyd et al 2013). This drop in the
density of FRII objects as the redshift decreases is captured in Figure 1, with
the smaller size of the FRII boxes. Fewer FRII quasars implies less hot halo
gas-accreting FRI radio galaxies, since, as noted above and as emphasized by the
arrows in the diagrams, the two are evolutionarily linked in the gap paradigm. 
What remains, therefore, are lower-mass accreting black holes, which precludes
both the existence of retrograde and hot mode accretion, forcing accretion to
exist in lower-mass black holes in prograde configurations, with the
radiatively-efficient subclass of such objects being radio quiet or without jets.
The non radiatively-efficient fraction, as can be seen on the diagrams, are
LINERs.  A bird's eye view of this suggests that compatibility with the Soltan
argument is well described as the result of an evolutionary process in which
accretion begins to dominate over jets, with an absence of retrograde systems and
an appearance of jets only in intermediate prograde spinning black holes in
radiatively efficient states. But the diagram captures the essential outcome of
prolonged accretion in the paradigm: rapidly spinning, dead black holes at low
redshift.

With the radio luminosity function as a proxy for jet power, and the
X-ray/optical luminosity function as a proxy for accretion power, this framework
is the first to make sense of a long-standing puzzle in the evolution of
extragalactic radio sources, i.e. the compatibility of both the radio and
X-ray/optical luminosity functions vs redshift. Within the context of the spin
paradigm, in fact, high spinning prograde accreting black holes produce the most
powerful jets, which observationally are detected in the radio band.  But
high-spinning prograde black holes have the highest disk efficiency if they are
thin disks.  And they are detected in the X-ray/optical band. Hence, according to
the simplest interpretation of the spin paradigm, the radio and X-ray/optical
luminosity functions should track each other. The observational inference,
instead, appears to be that accretion reaches its peak significantly later. Hence
the spin paradigm puzzle or `Meier paradox' (Meier 2012). We have proposed a
scenario that qualitatively resolves the conflict.

\section{Comparison to GRMHD}

In this section we wish to address some of the confusion that has emerged over
the differences between the ideas in the gap paradigm and the results of general
relativistic magnetohydrodynamic numerical simulations (GRMHD), which find that
prograde accreting black holes produce slightly more powerful jets than
retrograde ones (e.g. Tchekhovskoy \& McKinney 2012).  The take away point should
be that the semi-analytic foundation of the gap paradigm shares little common
ground with GRMHD so that differences should be expected.  The major differences
between GRMHD and the gap paradigm are twofold: 1) There is an uncertainty as to
what is and where the dynamo acts to produce the magnetic field and whether
accretion disks predominantly advect pre-existing fields or create them in-situ
(e.g. Blackman 2012 and references therein). GRMHD allows the accretion disk to
act as a dynamo; hence, the closer the disk is to the black hole, the greater the
field strength.  By this we simply mean that GRMHD equations naturally lead to
dynamo behavior while the gap paradigm forbids dynamo-like behavior in the
accretion disk, hence, the first fundamental difference. And such a difference is
largely responsible for determining whether retrograde or prograde black holes
have larger black hole-threading fields (Garofalo 2009). And the fact that
dynamo-like behavior naturally emerges from the GRMHD equations is true even in
the context of the recent magnetically dominated or flooded simulations where
there is an additional largescale magnetic field that is advected into the black
hole region by fiat.  In fact, this largescale magnetic field will be allowed to
thread the black hole or be diffused outward, depending on the field that threads
the disk, via magnetic field pressures.  But as field diffusion weakens the black
hole threading flux, the inner disk dynamo of the prograde accreting system will
enhance the magnetic field there, contributing additional magnetic pressure to
hold the field on the hole. Of course this begs the question of what mechanism
operates to provide the advected largescale field in the first place. While the
numerical solution presented in our scheme shares the unexplained assumption of a
pre-existing large-scale field, no dynamo-like behavior occurs in the accretion
disk, which is thus constrained to behave as a passive advector of magnetic flux,
allowing the retrograde system to experience the larger black hole threading flux
due to the advectively-prone larger gap region.  In other words, the no-flux
boundary condition in the gap region is responsible for the differences in black
hole-threading flux between the prograde and retrograde regimes. A real
astrophysical black hole system likely behaves in a way that lies in-between
these two extremes. 2) GRMHD adopts an ideal MHD scheme, ignoring the generalized
Ohm's law.  Much work has gone into showing that effective MHD parameters are
produced in turbulent regimes that do not require the specification of
microscopic physics (Guan \& Gammie 2009; Fromang \& Stone 2009; Lesur \&
Longaretti 2009; Eyink, Lazarian \& Vishniac 2011).  While there is evidence that
turbulent resistivity and diffusion appear to do a good job even in relativistic
regimes (Cho \& Lazarian 2014), there are issues that may require non-ideal MHD,
such as the generalization of the notion of flux-freezing in terms of
`magnetofluid connectivity' that may considerably alter the plasma behavior
(Asenjo \& Comisso 2015). These missing non-ideal terms that also ensure
causality via their time dependence, may become important in the violently
dynamical environment near black holes which would only be captured in the GRMHD
equations if the microscopic physics were specified. In fact, current MHD work is
attempting to bridge this gap in creative ways such as to change the spatial
interpolation for the hydrodynamic equations and magnetodynamics equations,
thereby mimicking a change in dissipative scales (Tchekhovskoy, personal
communication). While the gap paradigm includes the simplified Ohm's law, it is
quite old-school in this respect, adopting the $\alpha{}$-prescription. While two
issues - the origin of magnetic fields and the nature and limitation of
turbulent-driven, effective MHD in black hole plasmas -- constitute a gap in our
ultimate understating of strongly relativistic MHD systems that leave us with
doubts in the modeling of the spin dependence of jets in both analytic and
semi-analytic models like the gap paradigm as well as in numerical simulations of
black hole systems, we should not be surprised that the different approaches
produce different results. Ultimately, our position is that once GRMHD
simulations advance beyond the idealized Ohm's law, incorporating both radiation
and the required microphysics, the importance of retrograde accretion will
emerge.

\section{Summary and conclusion}

The gap paradigm for black hole accretion and jet formation constitutes a
phenomenological framework for the time evolution of AGN whose constraints from a
simple prolonged accretion picture predict a specific redshift distribution for a
large family of AGN.  While the powerful FRII quasars are modeled as retrograde
spinning systems, the implicit time evolution due to accretion ensures that their
black holes spin down toward zero spin and up into the prograde regime.  For the
accreting systems that remain in radiatively efficient states, only for
intermediate prograde black hole spins are jets allowed, as recent observations
suggest in NLS1s (Doi et al, ApJ, in press; Liu et al. 2015).  For high-spinning black holes in radiatively efficient accretion states,
the model prescribes jet suppression, a scale-invariant mechanism whose
by-products involve both jet power peak at higher redshift compared to accretion
peak -- qualitatively resolving the Meier Paradox - and radio quietness
associated with high spinning black holes at low redshift, as we also conclude
from the Soltan argument.  Our results are important because they constitute the
first and only application of the model to star formation, culminating in a
picture in which a subclass of the FRII quasars lead to an increase in the star
formation rate resulting from the fact that large gap regions between accretion
disks and black holes lead to both powerful and collimated jets as well as to
thermal disks with lower peak energy, allowing jet-induced star formation to
dominate over the negative feedback of disk winds. The simple time evolution of
the gap paradigm has now been applied to at least qualitatively explain a host of
seemingly different observations that in the model, instead, have a common
explanation: the radio loud/radio quiet dichotomy, the FRI/FRII break, the
difference in peak accretion power vs peak jet power vs redshift, the existence
of high-spinning radio quiet AGN at low redshift, the jet-disk connection, the
existence of more massive black holes in the radio loud quasar population, the
reason why jets occur in intermediate prograde spinning black holes in NLS1s, the
reason why disk winds are more powerful in radio quiet AGN/quasars compared to
radio loud AGN/quasars, and, finally most recently to the radio quasar-star
formation link. The remarkable number of observations that fit within the simple
evolutionary picture of the gap paradigm argues that retrograde accretion is an
essential element in our understanding of the cosmic evolution of black holes,
one that numerical models capable of including the physics of the central engine
in active galaxies should eventually incorporate.


\acknowledgements
DG thanks Caltech astrophysicist David L. Meier for identifying and explaining
the `Meier Paradox'.

\noindent {\bf REFERENCES}\\
Ansejo, F.A. \& Comisso, L. 2015, PRL, in print\\
Antonuccio-Delogu, V. \& Silk, J., 2010, MNRAS, 405, 1303\\
Barger, A.J. et al, 2014, ApJ, in press\\
Barger, A.J. et al, 2001, ApJ, 122, 2177\\
Bertone, S. \& Conselice, C.J., 2009, MNRAS, 396, 2345 \\
Best, P.N. \& Heckman, T.M. 2012, MNRAS, 421, 1569\\
Blackman, E.G., 2012, Physica Scripta, 86, 5\\
Blandford, R.D., 1990, 161B\\
Blandford, R.D. \& Znajek, R.L., 1977, MNRAS, 179, 433\\
Blandford, R.D. \& Payne, D.G., 1982, MNRAS, 199, 883\\
Bonfield D.G. et al 2011, MNRAS, 416, 13 \\
Bongiorno et al 2012, MNRAS, 427, 3103 \\
Brenneman, L. 2013, Acta Polytechnica, 53, 652\\
Cattaneo, A. et al 2009, Nature, 460, 213\\
Cattaneo, A., Dekel, A., Faber, S.M. \& Guiderdoni, B. 2008, MNRAS, 389,
567\\
Cho, J. \& Lazarian, A., 2014, ApJ, 780, 30\\
Cisternas, M. et al 2011, ApJ, 726, 57\\
Doi, A. et al, 2015, ApJ, 798, 30\\
Crockett, R. et al., 2012, MNRAS, 421, 1603\\
Croft, S. et al , 2006, AJ, 647, 1040\\
Draper, A.R. \& Ballantyne, D.R., 2012, ApJ, 751, 72\\
Dunlop, J.S., McLure, R.J., Kukula, M.J., Baum, S.A., O'Dea, C.P.,
Hughes, D.H., 2003, MNRAS, 340, 1095\\
Eyink, G.L., Lazarian, A. \& Vishniac, E. T., 2011, ApJ, 743, 51  \\
Fromang \& Stone, J., 2009, A\&A, 507, 19 \\
Feltre, A. , et al 2013, MNRAS, 434, 2426 \\
Ferrarese \& Merritt 2000, ApJ, 539, L9\\
Floyd,D. J.E., Dunlop, J.S., Kukula M.J., 2013, MNRAS, 429, 2\\
Fragile, C.P. \& Meier, D.L., 2009, ApJ, 693, 771\\
Fragile, C.P. Wilson, J. \& Rodriguez, M., 2012, MNRAS, 424, 524 \\
Gaibler, V. et al, 2012, MNRAS, 425, 438\\
Garofalo, D., Kim I. M., \& Christian, D.J., 2014, MNRAS, 442, 3097\\
Garofalo, D., 2009a, ApJ, 699, 400\\
Garofalo, D., 2009b, ApJ, 699L, 52\\
Garofalo, D., MNRAS, 2013(b), 434, 3196\\
Garofalo, D., AdvAstr. 2013(a), 213105\\
Garofalo, D., Evans, D.A., \& Sambruna, R.M., 2010, MNRAS, 406, 975 \\
Gebhardt et al 2000, ApJ, 539, L13\\
Guan \& Gammie, 2009, ApJ, 697, 1901\\
Harrison C.M. et al 2012, ApJ, 760, L15 \\
Hickox, R.C. et al 2014, ApJ, 782, 9\\
Juneau S. et al 2013, ApJ, 764, 176 \\
Kalfountzou, E. e tal 2014, MNRAS, 442, 1181\\
Kalfountzou E., Jarvis, M.J., Bonfield, D.G., Hardcastle, M.J., 2012,
MNRAS, 427, 2401\\
Kim, M.I., Christian, D.J., \& Garofalo, D., 2015, in preparation\\
Kormendy \& Richstone 1995, ARA\&A, 33, 581\\
Kuncic, Z \& Bicknell, G.V., 2004, ApJ, 616, 669\\
Kuncic, Z \& Bicknell, G.V., 2007, Ap\&SS, 311, 127\\
Lesur \& Longaretti, 2009, A\&A, 504, 309\\
Liu, Z., Yuan W., Lu Y., Zhou, X., 2015, MNRAS, 447, 517\\
Luts D. et al 2008, ApJ, 684, 853 \\
Magorrian et al 1998, AJ, 115, 2285\\
Marconi \& Hunt 2003, ApJ, 589, L21\\
McKinney, J.C., Tchekhovskoy, A., \& Blandford, R.D., 2012, MNRAS, 423,
3083\\
McLure, R.J. \& Jarvis, M.J., 2004, MNRAS, 352L, 45\\
Meier, D.L., Black Hole Astrophysics: The Engine Paradigm, Springer
Verlag Berlin Heidelberg, 2012\\
Meier, D.L., 2001, ApJ, 548, L9\\
Merloni A. et al 2010, ApJ, 708, 137\\
Moderski R., Sikora M., Lasota J-P, 1998, MNRAS, 301, 142\\
Moderski, R. \& Sikora M. 1996, A\&A, 120, 591\\
Neilsen, J. \& Lee, J.C., 2009, Nature, 458, 481\\
Page, M.J., Symeonidis, M., Vieira, J.D.,  et al. 2012, Nature, 485,
213\\
Penna, R.F. et al, 2010, MNRAS, 408, 752 \\
Perego, A. et al, 2009, MNRAS, 399, 2249 \\
Ponti, G. et al, 2012, MNRAS, 422,11\\
Pringle, J.E.,1981, ARA\&A, 19, 137\\
Rafferty D.A. et al 2011, ApJ, 742, 3 \\
Reynolds, C.S., 1997, MNRAS, 286, 513\\
Reynolds, M. T., Walton, D.J., Miller, J.M. \& Reis, C., 2014, ApJ, 792,
19 \\
Rosario D.J. et al 2012, A\&A, 545, A45 \\
Sambruna, R.M. et al, 2011, ApJ, 734, 105\\
Sikora, M. \& Begelman, M. 2013, ApJ, 764, 24\\
Singal, J. et al 2013, ApJ, 764, 43 \\
Shakura N.I. \& Syunyaev R.A., 1973, A\&A, 24, 337\\
Shao L. et al 2010, A\&A, 518, L26 \\
Sikora M., Stawarz, L., \& Lasota, J-P., 2007, ApJ, 658, 815\\
Soltan, A. 1982, MNRAS, 200, 115\\
Tadhunter, C. et al, 2014, Nature, 1038, 13520\\
Terlevich, R., Tenorio-Tagle, G., Franco, J., \& Melnick, J., 1992,
MNRAS, 255, 713 \\
Terlevich, R. \& Melnick, J. 1985, MNRAS, 213, 841\\
Tombesi, F. et al, 2010, A\&A, 521, A57(a)\\
Tombesi, F., Sambruna R.M. \& Reeves, J.N. et al, 2010, ApJ, 2010, 719,
700(b)\\
Trakhtenbrot, B. 2014, ApJ, 789, 9\\
Treister, E., Schawinski, K., Urry, C.M., Simmons, B.D. 2012, ApJ, 758,
39\\
Tremaine et al 2002, ApJ, 574, 740\\
Tchekhovskoy, A \& McKinney, J., 2012, MNRAS, 423L, 55\\
Van Velzen, S. \& Falcke, H., 2013, A\&A, 557, 7\\
Wilson, A.S., \& Colbert, E.J.M., 1995, ApJ, 438, 62\\
Wu, S., Lu, Y., Zhang, F. and Ye, L., 2013, MNRAS, 436, 327\\

\end{document}